\documentclass{elsart}
                                                                                
\usepackage{amssymb}
                                                                                
\begin{document}
                                                                                
\begin{frontmatter}
                                                                               
 \title{Landau gauge Jacobian and BRST symmetry}
 \author{
M. Ghiotti$^{a}$,
A.C. Kalloniatis$^{a}$,
A.G. Williams$^{a}$}
 \address{$^a$Centre for the Subatomic Structure of Matter,
University of Adelaide, South Australia 5000, Australia 
}
                                                                                

\begin{abstract}
We propose a generalisation of 
the Faddeev-Popov trick for Yang-Mills fields
in the Landau gauge.
The gauge-fixing is achieved as a genuine change
of variables.
In particular the Jacobian that appears 
is the modulus of the standard Faddeev-Popov determinant.  
We give a path integral representation of this in terms of 
auxiliary bosonic and Grassman fields extended beyond the usual
set for standard Landau gauge BRST. 
The gauge-fixing Lagrangian density 
appearing in this context is local and enjoys 
a new extended BRST and anti-BRST symmetry
though the gauge-fixing Lagrangian density in this case is not
BRST exact. 

\end{abstract}
                                                                                
\thanks{Preprint numbers: ADP-05-13/T623.}
\end{frontmatter}
                                                                                

\section{Introduction}                                                                                
The elevation of Faddeev-Popov (FP) gauge-fixing of
Yang-Mills theory beyond the
realm of perturbation theory has been intensely pursued in
recent years for many reasons.
Nonperturbative gauge-fixed calculations on the lattice
are being compared to analogous solutions of Schwinger-Dyson
equations \cite{Alkofer:2000,Bowman:2004}. 
As well, the long-term goal of simulating the
full Standard Model using lattice Monte Carlo requires
the Ward-Takahashi identities associated with BRST symmetry
\cite{Becchi:1975} in order to control the lattice renormalisation. 
The main impediment to nonperturbative gauge-fixing
is the famous Gribov ambiguity \cite{Gribov:1977}: 
gauges such as Landau and Coulomb
gauge do not yield unique representatives on gauge-orbits 
once large scale field fluctuations are permitted.  
To some extent one could live with such non-uniqueness
if one could incorporate all Gribov copies in a computation. However
the no-go theorem of Neuberger \cite{Neuberger:1986}
obstructs even this: (a naive generalisation
of) BRST symmetry forces a complete cancellation of all Gribov copies 
in BRST invariant observables giving $0/0$ for expectation values.
In particular, Gribov regions contribute with alternating
sign of the FP determinant.

Here we shall propose an approach which takes seriously
that gauge-fixing when seen as a change of variables involves
a Jacobian being the absolute value of the 
Faddeev-Popov determinant. Usually the absolute
value is dropped either because of an {\it a priori} restriction
to perturbation theory or because of the identification
of the determinant in terms of an invariant of a
topological quantum field theory \cite{Birmingham:1991} 
such as the Euler character \cite{Hirschfeld:1979,Baulieu:1996}. 
In the latter case the Neuberger problem is encountered.

The approach we describe in the following is not restricted to
perturbation theory. Moreover, because it will be seen to
involve a gauge-fixing Lagrangian density that is not
BRST exact it falls outside the scope of the preconditions
for the Neuberger problem. 
In the next section we shall derive the Jacobian associated
with gauge-fixing in the presence of Gribov copies. We shall
give a representation of the ``insertion of the identity'' in this
case in terms of a functional integral over an enlarged set
of scalar and ghost fields. The extended BRST symmetry of this
new gauge-fixing Lagrangian density will be described
though we will see that the final form of the gauge-fixing
Lagrangian is not BRST exact.

\section{Field theoretic representation for the Jacobian of FP gauge fixing}
In the following we shall formulate the problem in the
continuum approach to gauge theory.
  
Our aim is to generalise the standard formula from calculus
for a change of variable:

\begin{equation}
\label{change}
\left| \det\left({ {\partial f_i}\over {\partial x_j} }\right) 
\right|^{-1}_{{\vec f}=0}
= \int dx_1 \dots dx_n \delta^{(n)}({\vec f}({\vec x})).
\end{equation}

Here one is changing from integration variables ${\vec x}$ to
those satisfying the condition ${\vec f}({\vec x})=0$
and where, for Eq.(\ref{change}) to be valid,
in the domain of integration of $\vec x$ there 
must be only one such solution. 
In the context of gauge-fixing of Yang-Mills theory the
generalisation of Eq. (\ref{change}) is
\begin{equation}
\left| \det\left({ {\delta F[{}^gA]}\over {\delta g} }
\right) \right|^{-1}_{F=0}
= \int \mathcal{D}g \,\delta[F[{}^g\!A]] 
\label{functchange}
\end{equation}
where $A_{\mu}$ represents the gauge field, $g$ is an element
of the $SU(N)$ gauge group, 
$\mathcal{D}g$ is the functional integration measure in the group and 
\begin{equation}
F[{}^g\!A]=0
\label{gfcondition}
\end{equation}
is the gauge-fixing condition. We shall be interested 
in Landau gauge $F[A]=\partial_{\mu} A_{\mu}$. 
As in the calculus formula, here
Eq.(\ref{functchange}) is only valid as long as 
Eq.(\ref{gfcondition}) has a unique solution.
This is known not to be the case for Landau gauge. 
The FP operator nevertheless is
$M_F[A]= (\delta F[{}^gA]/\delta g)|_{F=0}$ and its determinant is
$\Delta_F[A]=\det(M_F)$.
For the Landau gauge 
$M_F[A]^{ab} = \partial_{\mu} D^{ab}_{\mu}[A]$ with $D^{ab}_{\mu}[A]$ the
covariant derivative with respect to $A_{\mu}^a$ in the 
adjoint representation.
Now the standard FP trick is
the insertion of unity in the measure of the
generating functional of Yang-Mills theory realised via
the identity (which follows from the above definitions):
\begin{equation}
1=\int \mathcal{D}g \Delta_F[{}^g\!A] \delta[F[{}^g\!A]].
\label{resolveunity}
\end{equation}

By analogy with standard calculus, in the presence of multiple solutions to the gauge-fixing condition
Eq.(\ref{resolveunity}) must be replaced by
\begin{equation}
\label{newidentity}
N_{F}[A] =\int \mathcal Dg\,\delta(F[{}^g\!A])\,\Big|\det
M_{F}[{}^g\!A]\Big|,
\end{equation}
where $N_{F}[A]$ is the number of different solutions for the
gauge-fixing condition $F[{}^g\!A]=0$ on the orbit
characterised by $A$, where $A$ is any configuration on the 
gauge orbit in question for which $\det M_F\neq 0$.

It is known that Landau gauge
has a fundamental modular region (FMR), namely a set of
unique representatives of every gauge orbit which is moreover
convex and bounded in every direction 
\cite{Dell'Antonio:1991xt,Semenov:1982}. 
The following discussion can be found in more
detail in \cite{vanBaal:1992}.
Denoted $\Lambda$, the FMR is defined as the
set of absolute minima of the functional $V_A[g]=\int d^4x ({}^gA)^2$
with respect to gauge transformations $g$. The stationary points of 
$V_A[g]$ are those $A_{\mu}$ satisfying the
Landau gauge condition. 
The boundary of the FMR, $\partial \Lambda$,
is the set of degenerate absolute minima
of $V_A[g]$. $\Lambda$ lies within
the Gribov region $\Omega_0$ where the FP operator
is positive definite. The Gribov region is comprised of all of the local minima of $V_A[g]$.
The boundary of $\Omega_0$,
the Gribov horizon $\partial\Omega_0$, is where the FP operator
$M_F$ (which corresponds to the second order variation
of $V_A[g]$ with respect to infinitesimal $g$) acquires zero modes. 
When the degenerate absolute minima of $\partial\Lambda$
coalesce, flat directions
develop and $M_F$ develops zero modes. Such orbits cross the
intersection of $\partial\Lambda$ and $\partial\Omega_0$. 
The interior of the fundamental modular region is a smooth
differentiable and everywhere convex manifold. Orbits crossing the boundary 
of the FMR on the other hand will cross that boundary again at
least once corresponding to the degenerate absolute
minima. 

Though, at present, there is no practical computational 
algorithm for constructing the FMR, it exists and we will make use of it 
for labelling orbits, i.e., $A_{\rm u}$ are defined to be configurations
in the FMR, $A_{\rm u}\in \Lambda$.
Since every orbit crosses the fundamental modular
region once we are guaranteed to have $N_{F}\geq 1$. 
In turn the ${}^gA_{\rm u}$
fulfilling the constraint of Eq. (\ref{gfcondition}) would be every other gauge copy
of $A_{\rm u}$ along its orbit. Eq.(\ref{newidentity})
is equal to the number of Gribov copies
on a given orbit, $N_{GC}=N_{F}-1$, except 
that copies lying on any of the Gribov horizons ($\Delta_{F}=0$) 
do not contribute to $N_{F}$. 
 
The finiteness of $N_F$ in the presence of a regularisation 
leading to a finite number of degrees of freedom 
(such as a lattice formulation) can be argued as follows.
Consider two neighboring Gribov copies corresponding to
a single orbit. If they contribute to $N_F$ they cannot
lie on the Gribov horizon. Therefore they do not
lie infinitesimally close to each other along a flat direction,
namely they have a finite separation. This is true then for all
copies on an orbit contributing to $N_F$: all copies contributing
to $N_F$ have a finite separation.  
But the $g$ which create the copies of $A_{\rm u}$
belong to $SU(N)$ which has a finite group volume. 
Thus for each space-time point there is a finite number of
such $g$.
We conclude then for a regularised formulation that $N_{F}$ is finite. 

Consider then the computation
of the expectation value of a gauge-invariant operator $O[A]$ over an ensemble
of gauge-field configurations ${ A_{\rm u} }$ which is this set of unique
representatives of gauge orbits discussed above.  

Note that for a gauge-invariant observable, it makes no difference whether $A_{\rm u}\in \Lambda$
or if the $A_{\rm u}$'s are any other unique representatives of the orbits.

The expectation value on these configurations
\begin{equation}
\langle O[A] \rangle = 
{   {\int {\mathcal D}A_{\rm u} O[A_{\rm u}] e^{-S_{YM}}   } 
  \over  {\int {\mathcal D}A_{\rm u} e^{-S_{YM}}   }  } 
\label{expecvalue}
\end{equation}    
is well-defined.
Since in any regularised formulation $N_F$ is a finite positive integer,
we can legitimately
use Eq.(\ref{newidentity}) to resolve the identity
analogous to the FP trick
and insert into the measure of
integration for an operator expectation value. We thus have
\begin{equation}
\langle O[A]\rangle=\frac{\int \mathcal DA_{\rm u}\frac{1}{N_{F}[A_{\rm u}]}
\int \mathcal Dg\,\delta(F[{}^g\!A])\,\Big|\det 
M_{F}[{}^g\!A]\Big|\,O[A]\,
e^{-S_{YM}[A]}} {\int \mathcal DA_{\rm u}\frac{1}{N_{F}[A_{\rm u}]}
\int \mathcal Dg\,\delta(F[{}^g\!A])\,\Big|\det M_{F}[{}^g\!A]\Big|
\,e^{-S_{YM}[A]}}.
\label{expectvalGC}
\end{equation}
We can now pass $N_{F}[A_{\rm u}]$ under the group
integration ${\mathcal D}g$ and combine the latter with ${\mathcal D}A_{\rm u}$
to obtain the full measure of all gauge fields ${\mathcal D}({}^gA_{\rm u})$
which we can write now as ${\mathcal D}A$.
$N_{F}$ is certainly gauge-invariant: it is a property
of the orbit itself. 
So $N_{F}[A_{\rm u}]=N_{F}[{}^gA_{\rm u}]
=N_{F}[A]$.
Thus we can write
\begin{equation}
\langle O[A]\rangle=\frac{\int \mathcal DA\,
\frac{1}{N_{F}[A]}
\delta(F[{}^g\!A])\,\Big|
\det M_{F}[{}^g\!A]\Big|\,O[A]\,e^{-S_{YM}[A]}}
{\int \mathcal DA\,
\frac{1}{N_{F}[A]}
\delta(F[{}^g\!A])\,\Big|\det
M_{F}[{}^g\!A]\Big|\,e^{-S_{YM}[A]}}.
\label{expecvaluewithcopies}
\end{equation}

Perturbation theory can be recovered from this of
course by observing that only $A$ fields near the trivial orbit, containing 
$A=0$ and for which $S_{YM}[A]=0$, contribute significantly in the perturbative regime: the curvature of the
orbits in this region is small so that the different orbits in
the vicinity of $A=0$ intersect the gauge-fixing hypersurface
$F=0$ the same number of times.
Then the number of Gribov copies is the same for each orbit, 
$N_{F}$ is independent of $A_{\rm u}$ and
we can cancel $N_{F}$ out of the expectation value. In that case
\begin{equation}
\label{expecvaluenocopies}
\langle O[A]\rangle=\frac{\int \mathcal DA\,\delta(F[A])\,\Big|
\det M_{F}[A]\Big|\,O[A]\,e^{-S_{YM}[A]}}
{\int \mathcal DA\,\delta(F[A])\,\Big|\det 
M_{F}[A]\Big|\,e^{-S_{YM}[A]}}.
\end{equation}       
In turn, observing that fluctuations near the trivial orbit
cannot change the sign of the determinant, the modulus can
also be dropped and one recovers the usual starting point
for a standard BRST invariant formulation of 
Landau gauge perturbation theory.
Note that perturbation theory is built on the gauge-fixing surface
in the neighbourhood of $A=0$, which for a gauge-invariant quantity
will be equivalent to averaging over the Gribov copies of $A=0$ as in Eq. (\ref{expecvaluenocopies}).
For the non-perturbative
regime, the orbit curvature increases significantly and
in general there is no reason to expect that $N_{F}$ would be the same for each
orbit. Moreover the determinant can change sign. 
  
Let us focus on the partition function appearing in 
Eq. (\ref{expecvaluewithcopies})
\begin{equation}
{\mathcal Z}_{\rm{gauge-fixed}} = \int {\mathcal D}A
\, N_{F}^{-1}[A]
\Big| \det(M_F[A]) \Big| \delta(F[A])\, e^{-S_{YM}}
\label{partitionfn1}
\end{equation}
The objective is to generalise the BRST formulation of 
Eq.(\ref{partitionfn1}) such that it is valid beyond perturbation theory
taking into account the modulus of the determinant.
We thus start with the following representation:
\begin{equation}
\Big|\det (M_F[A])\Big| = {\rm sgn}(\det (M_F[A]))\, \det (M_F[A])
\label{moddet}. 
\end{equation}
As mentioned, the factor ${\rm {det}}(M_F[A])$ in Eq.(\ref{moddet}) is  
represented as a functional integral via the usual Lie algebra valued 
ghost and anti-ghost fields
in the adjoint representation of $SU(N)$. 
Let us label these as $c^a, {\bar c}^a$. It is usual also 
(see for example \cite{Nakanishi:1990}) 
to introduce a Nakanishi-Lautrup auxiliary field $b^a$.
Thus the effective gauge-fixing Lagrangian density 
\begin{equation}
{\mathcal L}_{\rm{det}} =
-b^a \partial_{\mu} A^a_{\mu} + \frac{\xi}{2} b^a b^a
+ {\bar c}^a M_F^{ab}c^b
\label{usualLag}
\end{equation}
yields \cite{Nakanishi:1990} 
\begin{equation}
\lim_{\xi\rightarrow 0}
\int {\mathcal D}{\bar c}^a {\mathcal D}c^a {\mathcal D}b^a
e^{-\int d^4x {\mathcal L}_{\rm{det}} }
= \delta(F[A]) \det(M_F[A]). 
\end{equation}
In order to write the factor ${\rm {sgn}}(\det(M_F[A]))$   
in terms of a functional integral
weighted by a local action, we consider the following Lagrangian density
\begin{equation} 
\mathcal{L}_{{\rm{sgn}}} = i B^a M_F^{ab} {\varphi}^b - i {\bar d}^a M_F^{ab} d^b
+ \frac{1}{2} B^a B^b
\label{Lagsgn1}
\end{equation}
with ${\bar d}^a,d^a$ being new Lie algebra valued Grassmann fields 
and $\varphi^a,B^a$ being new auxiliary commuting fields.
Consider in Euclidean space the path integral 
\begin{equation}
\mathcal{Z}_{{\rm{sgn}}}= 
\int \mathcal{D}{\bar d}^a \mathcal{D}d^a \mathcal{D}\varphi^a \mathcal{D}B^a 
e^{-\int d^4 x \mathcal{L}_{{\rm{sgn}}}}\,.
\label{sgnpartition}
\end{equation}
Completing the square in the Lagrangian density of Eq.(\ref{Lagsgn1}), 
the $B$ field can be integrated out in the partition function
leaving an effective Lagrangian density
\begin{equation}
\mathcal{L}'_{{\rm{sgn}}} = \frac12 \varphi^a ((M_F)^T)^{ab} M_F^{bc} \varphi^c 
- i {\bar d}^a M_F^{ab} d^b\,,  
\label{sgnLag2}
\end{equation}
where $(M_F)^T$ denotes the transpose of the FP operator.
Integrating all remaining fields now 
it is straightforward to see that the partition function  
Eq.(\ref{sgnpartition}) amounts to just 
\begin{eqnarray}
\mathcal{Z}_{{\rm{sgn}}} &=& 
{ { {\rm{det}}(M_F)} \over {\sqrt{{\rm{det}}((M_F)^T M_F)}} } = {\rm{sgn}} ({\rm{det}}(M_F))\,. 
\end{eqnarray}
Thus the representation Eq.(\ref{sgnpartition}) can be used
for the first factor of Eq.(\ref{moddet}). The Lagrangian density of 
Eq.(\ref{Lagsgn1}) therefore combines with   
the standard BRST structures of Eq.(\ref{usualLag})
coming from the determinant itself in Eq.(\ref{moddet})
so that an equivalent representation for the partition function
based on Eq.(\ref{partitionfn1}) is
\begin{equation}
{\mathcal Z}_{\rm{gauge-fixed}} =
\int {\mathcal D}A^a_{\mu} {\mathcal D}{\bar c}^a {\mathcal D}c^a
{\mathcal D}{\bar d}^a {\mathcal D}d^a {\mathcal D}b^a {\mathcal D}\varphi^a
(N_{F}[A])^{-1}\, e^{-S_{\rm YM}-S_{\rm \det}-S_{\rm sgn}}
\end{equation}
with $S_{\det}$ and $S_{\rm sgn}$ the actions corresponding to
the above Lagrangian densities Eqs. (\ref{usualLag},\ref{Lagsgn1}).
 
\section{A new extended BRST}

The symmetries of the new Lagrangian density, $\mathcal L_{\rm sgn}$, 
are essentially a boson-fermion
supersymmetry and can be seen from Eq.(\ref{Lagsgn1}). 
In analogy to the standard BRST transformations typically denoted by $s$,
we shall denote them by the Grassmann graded operator $t$
\begin{eqnarray}
t \varphi^a &=& d^a \nonumber \\
t d^a &=& 0 \nonumber \\
t {\bar d}^a &=& B^a \nonumber \\
t B^a &=& 0\,,  
\label{t-alg}
\end{eqnarray}
such that
\begin{equation}
t\mathcal L_{\rm sgn}=0
\end{equation}
and trivially $t\mathcal L_{\rm YM}=0$. 
Eqs.(\ref{t-alg}) realise the infinitesimal form of shifts in
the fields. The operation
$t$ is nilpotent: $t^2=0$. Using Eqs.(\ref{t-alg}) we can
give the following form for the Lagrangian density $\mathcal{L}_{\rm{sgn}}$,
\begin{equation}
\mathcal{L}_{\rm{sgn}}= 
t \left( {\bar d}^a (i M_F^{ab} \varphi^b + \frac{1}{2} B^a) \right) .
\end{equation}
The question now is how to combine this with the standard BRST
transformations 
\begin{eqnarray}
s A_{\mu}^a &=& D_{\mu}^{ab} c^b \nonumber \\
s c^a &=& - \frac12 g f^{abc} c^b c^c \nonumber \\
s{\bar c}^a &=& b^a \nonumber \\
s b^a &=& 0 .
\end{eqnarray}
The transformations due to $t$ and $s$ are completely
decoupled except that the latter also act on the gauge field
on which the FP operator $M_F$ depends. We propose the
following unification of these symmetry operations.
Consider an operation $\mathcal{S}$ {\it block-diagonal}
in $s$ and $t$: $\mathcal{S}={\rm{diag}}(s,t)$.
The operator acts on the following multiplet fields:
\begin{eqnarray}
\mathcal{A}^a = 
\left( \begin{array}{c} A^a_{\mu} \\\varphi^a \end{array} \right), \
\mathcal{C}^a= 
\left( \begin{array}{c} c^a \\d^a \end{array} \right), \ 
{\bar \mathcal{C}}^a= 
\left( \begin{array}{c} {\bar c}^a \\{\bar d}^a \end{array} \right), \ 
\mathcal{B}^a=
\left( \begin{array}{c} b^a \\B^a \end{array} \right).
\end{eqnarray} 
We see that these fields transform under $\mathcal{S}$ completely
analogously to the standard BRST operations
\begin{eqnarray}
\label{newalgebra}
\mathcal{S} \mathcal{A}^a &=& \mathcal{D}^{ab} \mathcal{C}^b \nonumber \\  
\mathcal{S} \mathcal{C}^a_i &=& 
\mathcal{F}^{abc}_{ijk} \mathcal{C}^b_j \mathcal{C}^c_k 
\nonumber \\
\mathcal{S} {\bar \mathcal{C}}^a &=& \mathcal{B}^a \nonumber \\
\mathcal{S} \mathcal{B}^a &=& 0\,, 
\end{eqnarray}
where $i,j,k=1,2$ label the elements of the multiplets, and  

\begin{eqnarray}
 \mathcal{D}^{ab} &=& {\rm{diag}} (D_{\mu}^{ab}, \delta^{ab}) \nonumber \\
\mathcal{F}^{abc}_{111} &=& -\frac12 g f^{abc} ,\quad \mathcal F_{ijk}^{abc}=0 \quad {\rm for} \quad ijk\neq 111.  
\end{eqnarray}

Note that nilpotency is satisfied, $\mathcal{S}^2=0$. 
We shall refer to this type of operation as an {\it extended} BRST 
transformation which we distinguish from the BRST--anti-BRST
or double BRST algebra of the Curci-Ferrari model
\cite{Curci:1976,Thierry-Mieg:1985}.
We can thus formulate the gauge-fixing Lagrangian density  
for the Landau gauge as 
\begin{equation}
\mathcal{L}_{\rm gf} = {\rm Tr} \ \mathcal{S} 
\left( \begin{array}{cc}
{\bar c}^a F^a & 0 \\
0 & {\bar d}^a(i M_F^{ab} \varphi^b + \frac{1}{2} B^a) \end{array} \right) 
.
\label{gfLag}
\end{equation}

This approach admits also an extended anti-BRST operation:

\begin{eqnarray}
{\bar {\mathcal{S}}} \mathcal{A}^a &=& 
\mathcal{D}^{ab} {\bar {\mathcal{C}}}^b \nonumber \\
{\bar {\mathcal{S}}} {\bar \mathcal{C}}^a_i &=& \mathcal{F}^{abc}_{ijk} 
{\bar {\mathcal{C}}}^b_j {\bar {\mathcal{C}}}^c_k \nonumber \\
{\bar {\mathcal{S}}} \mathcal{C}^a &=& -\mathcal{B}^a \nonumber \\
{\bar {\mathcal{S}}} \mathcal{B}^a &=& 0\,.
\end{eqnarray}

Writing ${\bar \mathcal{S}} = {\rm{diag}} ({\bar s}, {\bar t})$
we can extract the standard anti-BRST ${\bar s}$-operations 
\cite{Thierry-Mieg:1985,Baulieu:1981} 
\begin{eqnarray}
{\bar s} A_{\mu}^a &=& D_{\mu}^{ab} {\bar c}^b \nonumber \\
{\bar s} {\bar c}^a &=& - \frac12 g f^{abc} {\bar c}^b {\bar c}^c \nonumber \\
{\bar s} c^a &=& -b^a \nonumber \\
s b^a &=& 0 
\end{eqnarray}
and those corresponding to ${\bar t}$:

\begin{eqnarray}
{\bar t} \varphi^a &=& {\bar d}^a \nonumber \\
{\bar t} {\bar d}^a &=& 0 \nonumber \\
{\bar t} d^a &=&-B^a \nonumber \\
{\bar t} B^a &=& 0\,.
\end{eqnarray}

Moreover, the ghosts and anti-ghosts in this extended structure
also fulfill the criteria for being Maurer-Cartan one-forms,

\begin{equation}
\mathcal{S} {\bar \mathcal{C}} + {\bar \mathcal{S}} \mathcal{C} =0
.
\end{equation}

However there is no extended 
BRST--anti-BRST (or double) symmetric
form of the gauge-fixing Lagrangian density Eq. (\ref{gfLag}),
unlike the two pieces of which it consists.
Such a representation exists
in the $s-$sector of Landau gauge:

\begin{equation}
\mathcal{L}_{{\rm gf}, s} = \frac12s {\bar s} A^a_{\mu} A^a_{\mu}\,.
\end{equation}

In the $t-$sector, the corresponding structure is

\begin{equation}
\mathcal{L}_{{\rm gf},t} = \frac12t {\bar t} 
\left [ \varphi^a M_F^{ab} \varphi^b +  {\bar d}^a d^a
\right] .
\end{equation}

However the complete Landau gauge-fixing Lagrangian density
can only be expressed via a trace, namely as 
\begin{equation}
{\mathcal L}_{\rm{gf}}= \frac12 {\rm Tr}
\mathcal{S}{\bar \mathcal{S}} \mathcal{W}
\end{equation}
with
\begin{equation}
     \mathcal{W} = {\rm{diag}} 
\left( A^a_{\mu} A^a_{\mu}, 
\varphi^a M_F^{ab} \varphi^b +  {\bar d}^a d^a 
\right)\,.
\end{equation}
Nevertheless this 
compact representation formulates the modulus of the determinant
in Landau gauge fixing in terms of a local Lagrangian density
and follows as closely as possible the standard BRST
formulation without the modulus.

\section{Discussion and Conclusions}
We have thus found a representation for Landau gauge-fixing
corresponding to the FP trick being an actual change of variables
with appropriate determinant. The resulting gauge-fixing Lagrangian
density enjoys a larger extended BRST and anti-BRST symmetry. However
it cannot be represented rigorously as a BRST exact object, rather
the sum of two such objects corresponding to different BRST operations.
This means that some of the BRST machinery is not available
to this formulation, such as the Kugo-Ojima criterion for 
selecting physical states. 
We discuss cursorily
now the perturbative renormalisability of the present formulation
of the theory. Note that the procedure leading to Eq. (\ref{gfLag})
does not introduce any new coupling constants; only the
strong coupling constant $g$ is present in $M_F[A]$ coupling
the Yang-Mills field to both the new ghosts and scalars.
The dimensions of the new fields are 

\begin{equation}
[\varphi]=L^0, \quad [d]=[\bar d] = L^{-1}, \quad [B]= L^{-2} .
\end{equation}

Most importantly in this context, the kinetic term for the
new boson fields $\varphi^a$ is {\it quartic} in derivatives:

\begin{equation}
\mathcal{L}_{\rm kin} = \varphi^a (\partial^2)^2 \varphi^a\,,
\end{equation}

which is renormalisable, by power counting, since $\varphi^a$ are dimensionless.
Such a contribution is seemingly harmless in the ultraviolet regime: for 
large momenta propagators will vanish like $1/p^4$. 
Moreover it should play an important role in guaranteeing
the decoupling of such contributions in perturbative diagrams.
That such a decoupling should occur is clear from Eq. (\ref{moddet}): in
the perturbative regime fluctuations about $A_{\mu}=0$ will not feel the 
${\rm {sgn}}({\rm {det}}M_F[A])$,
so that the field theory constructed in this way must
be equivalent to the perturbatively renormalisable Landau gauge fixed
theory. For example in the computation of the running
coupling constant we expect that this property will lead to
a complete decoupling of the $t$-degrees of freedom so that the
known Landau gauge result emerges from just the gluon and standard
ghost sectors.  
Naturally, the new degrees of freedom will be relevant in
the infra-red regime, which will be the object of future study.

\section*{Acknowledgements}
ACK is supported by the Australian Research Council.
We are indebted to discussions with Lorenz von Smekal, 
Mathai Varghese, Max Lohe and Martin Schaden.
                                                                                
                                                                                

\end{document}